\documentclass[sigconf]{acmart}

\usepackage{booktabs}
\usepackage{graphicx}
\usepackage{amsmath}
\usepackage{multirow}
\usepackage{xspace}
\usepackage{threeparttable}
\newcommand{\sys}{NeuroPriv\xspace}
\newcommand{\ba}{\mathrm{BA}}

\title{\sys: Adversarial Representation Learning for Privacy in Wearable EEG Systems}

\author{Sarmistha Sarna Gomasta}
\orcid{0000-0003-1618-4708}
\affiliation{%
  \institution{University of Massachusetts Amherst}
  \country{United States}}
\email{sgomasta@umass.edu}

\author{Bhawana Chhaglani}
\orcid{0000-0002-4060-4883}
\affiliation{%
  \institution{University of Massachusetts Amherst}
  \country{United States}}
\email{bchhaglani@umass.edu}

\author{Prashant Shenoy}
\orcid{0000-0002-5435-1901}
\affiliation{%
  \institution{University of Massachusetts Amherst}
  \country{United States}}
\email{shenoy@cs.umass.edu}

\copyrightyear{2026}
\acmYear{2026}
\setcopyright{cc}
\setcctype{by}

\acmConference[MobiDx '26]
{Hot Topics in Mobile and Wireless Systems for Health}
{October 26--30, 2026}
{Austin, TX, USA}

\acmBooktitle{Hot Topics in Mobile and Wireless Systems for Health
(MobiDx '26), October 26--30, 2026, Austin, TX, USA}

\acmDOI{10.1145/3842657.3842784}

\acmISBN{979-8-4007-2974-4/2026/10}
\begin{document}
\begin{abstract}
Wearable EEG systems may expose sensitive information beyond their intended health function, creating substantial risks to neuroprivacy. In this work, we show that commonly used EEG features can reveal participant identity and demographic attributes in addition to supporting the intended cognitive task. Wearable EEG is increasingly being explored for cognitive monitoring, neurological assessment, and longitudinal digital-health applications, yet many systems assume that transmitting compact spectral or spatial features instead of raw EEG provides sufficient privacy protection. Using EEGMAT as a motivating case study, we find that compact EEG features achieve a balanced accuracy of 0.788 for cognitive-state classification while enabling gender, age, and subject-identity inference with balanced accuracies of 0.858, 0.789, and 0.692, respectively. We further show that privacy-aware representation learning preserves task performance at 0.781 while reducing these inference accuracies to 0.563, 0.467, and 0.206. These findings motivate purpose-limited representations and explicit privacy auditing in wearable neurohealth systems.
\end{abstract}


\maketitle
\raggedbottom
\section{Introduction}
Wearable EEG sensing is moving beyond clinical and laboratory settings
into everyday computing \cite{nguyen2023comprehensive,duvinage2013performance}. Emerging head-worn systems support applications
such as cognitive-state monitoring, workload estimation, brain--computer
interaction, sleep analysis, and mobile health assessment \cite{berka2007eeg,mcfarland2011brain,arnal2020dreem,de2024sleep}. To make these applications practical on resource-constrained devices, wearable EEG
pipelines commonly replace raw-signal transmission with compact spectral and spatial
features that can be processed efficiently by downstream machine learning models.
This low-latency design remains privacy-sensitive. Traditional EEG features are often
treated as safer than raw signals because they remove waveform-level detail and explicit
identifiers. However, removing raw data does not necessarily remove identifiability
\cite{campisi2014brain}. EEG contains subject-specific and demographic information, and prior
work has shown that users can be identified from EEG-derived representations even when
direct identifiers are absent \cite{meng2023user}. In this work, we show that a wearable system
may avoid releasing raw EEG while still exposing sensitive information through the
features it transmits.

We examine this problem on EEGMAT \cite{zyma2019electroencephalograms}, where the intended utility task is
Baseline-versus-Arithmetic classification and the private attributes are
gender, age, and subject identity.Our analysis shows that the same feature representation used for task
inference also supports strong private-attribute inference. The best task
classifier achieves a balanced accuracy of 0.788, while gender, age, and
subject identity reach 0.858, 0.789, and 0.692, respectively. These
results indicate that compact feature release is not inherently
privacy-preserving and that the released representation itself must be
treated as sensitive.

We further analyze where this leakage originates. ANOVA-based feature
responsibility analysis shows that the private attributes are encoded
through different spectral and spatial feature groups. Gender leakage is
most strongly associated with theta power, age with beta-relative power
and asymmetry features, and subject identity with beta power and
alpha--beta relationships. This attribute-specific structure limits the
effectiveness of simple feature-level defenses. Uniform perturbation can
damage task-relevant information together with private information, while
fixed feature masking does not consistently suppress all private
attributes.

To address this limitation, we present \sys, a learned representation
framework for privacy-preserving wearable EEG feature release.
Figure~\ref{fig:architecture} shows the system pipeline. \sys operates
between feature extraction and downstream inference. It maps compact EEG
features and their pairwise interactions into a task-relevant latent
representation, then uses adversarial learning to reduce information
associated with gender, age, and subject identity before the
representation is released. The transformed representation is used for
the intended task, while external attackers are trained independently to
measure residual privacy leakage. We make the following contributions:

\begin{itemize}
    \item We demonstrate that compact EEG features leak gender, age, and
    subject identity even when raw EEG signals are not transmitted.
    
    \item We characterize the feature-level sources of privacy leakage
    and show that different private attributes are associated with
    different spectral and spatial feature groups.
    
    \item We evaluate Gaussian noise and feature masking as lightweight
    feature-level defenses and identify their limitations for
    multi-attribute privacy preservation.
    
    \item We design \sys, a learned representation framework that
    preserves task utility while suppressing gender, age, and subject
    identity information before feature release.
\end{itemize}
\section{Related Work}
EEG contains information beyond the signal component required by an intended
application. Early work demonstrated that consumer EEG devices can be exploited
as side channels to infer sensitive information from neural
responses~\cite{martinovic2012feasibility}. Subsequent studies showed that EEG
also contains persistent subject-specific patterns that support biometric
identification and cross-session re-identification~\cite{maiorana2015permanence,meng2023user}. Cryptographic approaches use secure computation to prevent direct disclosure of
EEG signals~\cite{agarwal2019protecting}. Federated and transfer learning
approaches keep recordings local while sharing model parameters or pretrained
models~\cite{jia2024federated,wei2025sandwich}. Synthetic EEG generation
supports downstream training without distributing the original patient
recordings~\cite{pascual2020epilepsygan}. These approaches protect EEG during
storage, sharing, or model training. They do not prevent sensitive attributes
from being inferred from a feature vector or embedding that a system releases
by design at inference time.

A parallel line of work partitions a network between a client device and a
server, so the server never observes the raw input. Abuadbba et al. showed that
intermediate activations of a 1D CNN over physiological time series remain
highly correlated with the raw signal, which undermines the assumption that a
split point alone confers privacy~\cite{abuadbba2020can}. Aziz et al. instantiate split inference on multi-channel
EEG headwear for seizure detection~\cite{aziz2024hardware}. They introduce a
Combined Privacy Metric that averages six normalized statistical, geometric,
and information-theoretic quantities computed between the raw input and the
intermediate activation, and they report that deeper convolutional split points
yield higher combined privacy scores at increasing client-side cost.

Two properties of that formulation motivate our design. First, each component
of the metric measures similarity between the released tensor and the raw
signal. Low similarity does not imply that a private attribute is hard to
recover, because an adversary is free to learn a nonlinear decision boundary in
the released space. Second, each component is normalized by its maximum across
the split points under evaluation, so the resulting score is relative to the
configurations tested rather than an absolute leakage quantity. \sys therefore
measures leakage operationally. We train independent external attackers on the
released representation and report their accuracy against a chance baseline.

Recent work applies perturbation and representation learning to reduce private
information while preserving EEG task utility \cite{meng2023user}. Prior work extends adversarial
perturbation to multiple private attributes, including identity, gender, and
BCI experience~\cite{meng2024protecting}. Autoencoder-based anonymization
and identity-removal networks similarly separate task-relevant information from
subject-specific information~\cite{singh2023selective,wang2025id}. These
studies demonstrate the value of learned privacy--utility optimization. They
operate on raw EEG, reconstructed EEG, or task-specific signal perturbations.

Our method, \textbf{\sys}, addresses privacy at the compact-feature release boundary of a wearable
EEG pipeline. We do not reconstruct or perturb the raw signal. We transform
spectral and spatial EEG features into a latent representation before release.
We quantify leakage for gender, age group, and subject identity using
independently trained external attackers, identify the feature groups
associated with each private attribute, and compare the learned representation
against Gaussian perturbation and privacy-ranked feature masking. This lets us
evaluate directly whether a released EEG representation preserves its intended
utility while reducing several distinct forms of private-attribute inference.

\section{Methods}

\subsection{Dataset and Features}

We use the EEGMAT dataset \cite{zyma2019electroencephalograms}, containing EEG recordings (36 people) collected
before and during the performance of mental arithmetic tasks. Each
subject contributes recordings under Baseline and Arithmetic conditions,
and the utility target is to distinguish these two states. In addition to
the task label, each sample is associated with subject-level private
attributes: gender, age group, and subject identity. 

Each EEG window is represented by a 17-dimensional handcrafted feature
vector. The representation includes absolute band powers, 
relative band powers, spectral ratios, and  spatial asymmetry
measures. The frequency bands are delta, theta, alpha, beta, and gamma.
The ratio features are theta/alpha, alpha/beta, theta/beta, and
engagement. The asymmetry features are left--right alpha asymmetry,
left--right beta asymmetry, and temporal--frontal alpha asymmetry.

\subsection{Threat Model and Leakage Metrics}

We consider a wearable EEG pipeline in which raw EEG remains on the device
and only a derived representation is released for downstream inference.
Depending on the system configuration, this released representation
$\mathbf{z}$ may correspond to the original compact EEG features, a
feature-level defended representation, or the latent representation produced
by \sys. We treat $\mathbf{z}$ as the privacy boundary of the system: data
upstream of $\mathbf{z}$ is protected by the device, whereas downstream
services that operate on $\mathbf{z}$ may be untrusted or semi-trusted.

Each EEG window is represented by $\mathbf{x}\in\mathbb{R}^{17}$ and is
associated with one utility label $y_T$ and three private labels: gender
$y_G$, age group $y_A$, and subject identity $y_I$. The intended utility task
is Baseline-versus-Arithmetic classification. We assume that any service or
third party with access to $\mathbf{z}$ may additionally attempt to infer
these private attributes using auxiliary labeled data from the same
population. The adversary does not observe the raw EEG, the parameters of the
\sys encoder, or the training gradients.

In many application scenarios, these attributes are sensitive. Subject
identity turns EEG into a biometric, enabling persistent re-identification and
cross-service tracking even when other identifiers are removed \cite{maiorana2015permanence,meng2023user}. Demographic
attributes such as gender and age support profiling and differential
treatment, and can be combined with other data sources to link neural signals
to medical or behavioral records \cite{douglas2026evaluating}. We therefore treat successful prediction of
$y_G$, $y_A$, or $y_I$ from $\mathbf{z}$ as privacy leakage, even though the
raw EEG never leaves the device and the representation is optimized for
low-latency inference.

Utility is measured using task balanced accuracy:
\begin{equation}
\label{eq:utility}
U(\mathbf{z}) =
\ba\big(y_T, h_T(\mathbf{z})\big),
\end{equation}
where $h_T$ is the downstream task classifier. Higher $U(\mathbf{z})$
indicates better preservation of task performance. Privacy leakage is measured using an attacker set
$\mathcal{A}=\{\mathrm{SVM},\mathrm{RF},\mathrm{GB}\}$. For each
private attribute $p\in\{G,A,I\}$, we report the strongest attack performance:
\begin{equation}
\label{eq:leakage}
P_p(\mathbf{z}) =
\max_{a\in\mathcal{A}}
\ba\big(y_p, a(\mathbf{z})\big).
\end{equation}
Here, $P_p(\mathbf{z})$ captures the best balanced accuracy an attacker can
achieve on attribute $p$ given only $\mathbf{z}$ and auxiliary labeled data.
Values near chance indicate limited recoverability of the corresponding
attribute, whereas substantial gains over chance reflect significant leakage.
The balanced-accuracy chance levels are $0.500$ for the binary task and gender
labels, $1/3=0.333$ for the three-class age label, and $1/36=0.028$ for
closed-set subject identity. We apply this leakage evaluation to the original
feature representation, to Gaussian-noise and feature-masking baselines, and
to the representation produced by \sys.

To identify which EEG features contribute most strongly to leakage, we compute
a one-way ANOVA $F$-value for each feature and private attribute. For feature
$j$ and private attribute $p$,
\begin{equation}
\label{eq:anova}
F_{j,p} =
\frac{
\sum_{c} n_c
\left(\bar{x}_{j,c}-\bar{x}_{j}\right)^2/(C-1)
}{
\sum_{c}\sum_{i\in c}
\left(x_{i,j}-\bar{x}_{j,c}\right)^2/(N-C)
}.
\end{equation}
Here, $C$ is the number of classes, $n_c$ is the number of samples in class
$c$, $\bar{x}_{j,c}$ is the class-specific mean of feature $j$, and
$\bar{x}_{j}$ is its overall mean. A larger $F$-value indicates stronger class
separability and therefore greater potential leakage for the corresponding
private attribute. We use these scores both to analyze feature-level leakage
and to construct the feature-masking baseline.

\subsection{Gaussian Noise}

Gaussian noise is a coordinate-agnostic privacy baseline. It perturbs
every standardized feature coordinate:
\begin{equation}
\tilde{\mathbf{x}} = \mathbf{x} + \boldsymbol{\epsilon},
\qquad
\boldsymbol{\epsilon} \sim \mathcal{N}(0, \sigma^2 \mathbf{I}).
\end{equation}
The defense uses no information about which features leak, so it applies
equal perturbation to attribute-bearing and task-bearing coordinates. We
sweep $\sigma$ and audit task and identity at each level.

\subsection{Feature Masking}

Feature masking is a coordinate-targeted baseline. We rank EEG
features using their combined contribution to gender, age, and subject
identity leakage and zero the top-$k$ privacy-ranked coordinates.

Let $R_j$ denote the combined privacy-responsibility score of feature
$j$ across gender, age, and subject identity. The masking transformation
is

\begin{equation}
M^{(k)}(\mathbf{x})_j =
\begin{cases}
0, & \text{if } \mathrm{rank}(R_j) \leq k, \\
x_j, & \text{otherwise}.
\end{cases}
\end{equation}

Because the features are standardized, setting a coordinate to zero
replaces it with its training-set mean.

\subsection{\sys Architecture}

\begin{figure*}[t]
\centering
\includegraphics[width=0.7\textwidth]{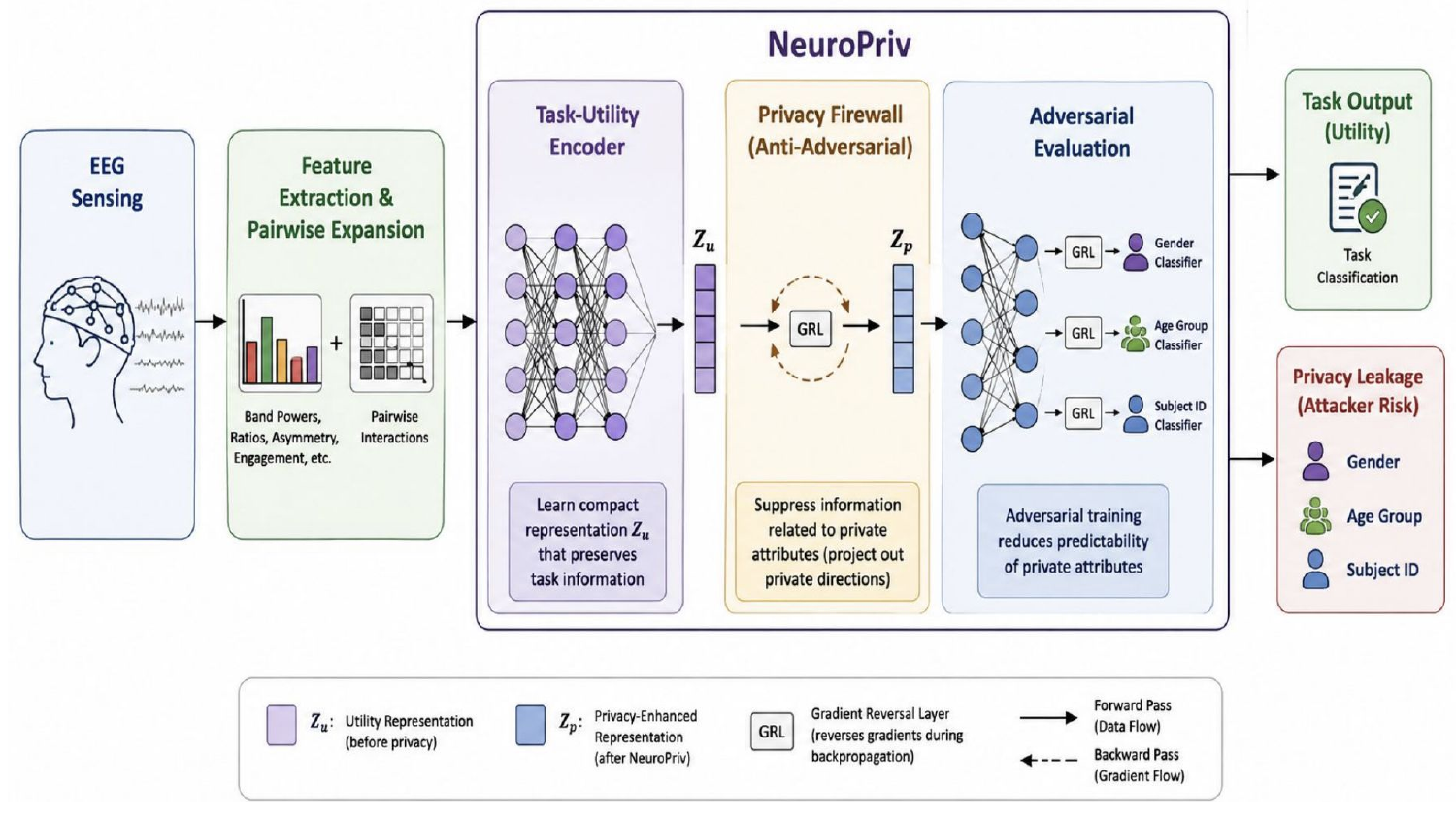}
\caption{\sys architecture. EEG features and pairwise
interactions are mapped to a task-relevant latent representation.
Gradient-reversal adversaries suppress gender, age, and subject identity
information before the representation is released for downstream
inference.}
\label{fig:architecture}
\end{figure*}

\sys transforms compact EEG features into a task-relevant representation
with reduced private-attribute information. As shown in
Figure~\ref{fig:architecture}, the framework consists of feature
expansion, latent representation learning, and adversarial privacy
suppression.

Given an EEG feature vector
$\mathbf{x}\in\mathbb{R}^{17}$, we augment the original features with
all pairwise interactions:

\begin{equation}
\phi(\mathbf{x})
=
\left[
\mathbf{x};
\left\{x_i x_j\right\}_{i<j}
\right]
\in\mathbb{R}^{153}.
\end{equation}

The expanded representation is mapped to a compact latent representation
through an encoder:

\begin{equation}
\mathbf{z}=f_{\theta}\big(\phi(\mathbf{x})\big).
\end{equation}

A task head predicts the intended utility label from $\mathbf{z}$,
while three privacy heads predict gender, age group, and subject
identity. The privacy heads are connected to the encoder through
gradient-reversal layers. During training, the task objective encourages
$\mathbf{z}$ to retain task-relevant information, whereas the reversed
privacy gradients discourage the encoder from preserving information
useful for private-attribute inference.

The resulting optimization objective is

\begin{equation}
\min_{\theta,h_T}
\max_{h_G,h_A,h_I}
\mathcal{L}_{T}
-
\lambda_G\mathcal{L}_{G}
-
\lambda_A\mathcal{L}_{A}
-
\lambda_I\mathcal{L}_{I},
\end{equation}

where $\mathcal{L}_{T}$ is the task-classification loss and
$\mathcal{L}_{G}$, $\mathcal{L}_{A}$, and $\mathcal{L}_{I}$ are the
gender, age, and identity losses.

After training, the privacy heads are removed and only the transformed
representation $\mathbf{z}$ is released for downstream task inference.
Privacy is evaluated independently using external SVM, RF, and GB
attackers trained on the released representation.

\section{Evaluation}

We evaluate four aspects of the proposed framework: 
(i) utility and privacy leakage in the original EEG feature
representation, 
(ii) the EEG features most responsible for private-attribute leakage,
(iii) Gaussian noise and feature masking as lightweight privacy
baselines, and 
(iv) the privacy--utility performance of \sys.

Balanced accuracy (BA) is used for both utility and privacy evaluation.
For the original feature representation, we report the performance of
support vector machine (SVM), random forest (RF), and gradient boosting
(GB) classifiers. For each private attribute, leakage is determined by
the strongest evaluated attacker, following Eq.~\ref{eq:leakage}.The feature-level defense experiments and the \sys experiment were
conducted using different data partitions. We therefore evaluate each
method relative to the corresponding original-feature baseline from the
same experimental protocol and do not directly compare absolute values
across the two protocols.
\paragraph{\textbf{Leakage in the original EEG representation}.}
Table~\ref{tab:baseline_all} reports task and private-attribute balanced
accuracy for the original EEG feature representation. RF achieves the
highest task balanced accuracy of 0.788. However, the same representation
also supports gender, age, and subject-identity inference with balanced
accuracies of 0.858, 0.789, and 0.692, respectively.
Gender is inferred more accurately than the intended cognitive task.
Subject-identity inference also substantially exceeds its closed-set
chance level of $1/36=0.028$. These results show that compact EEG
features preserve useful task information while simultaneously exposing
multiple private attributes.

\begin{table}[t]
\centering
\caption{Balanced accuracy for the original EEG feature representation
under each attacker.}
\label{tab:baseline_all}
\begin{tabular}{lccc}
\toprule
Target & SVM & RF & GB \\
\midrule
Task      & 0.639 & \textbf{0.788} & 0.763 \\
Gender    & 0.755 & 0.845 & \textbf{0.858} \\
Age       & 0.657 & \textbf{0.789} & \textbf{0.789} \\
SubjectID & 0.635 & \textbf{0.692} & 0.632 \\
\bottomrule
\end{tabular}
\end{table}

\paragraph{\textbf{Feature responsibility for privacy leakage}.}

Table~\ref{tab:feature_responsibility} reports the highest-ranked EEG
features according to their ANOVA $F$-values. The rankings differ across
private attributes, indicating that privacy leakage is distributed across
different spectral and spatial feature groups.

Gender is most strongly associated with theta absolute power
($F=75.135$) and theta relative power ($F=53.203$). Age is most strongly
associated with beta relative power ($F=42.683$) and temporal--frontal
alpha asymmetry ($F=41.877$). Subject identity is dominated by beta
absolute power, which reaches an $F$-value of 116.644, followed by beta
relative power and the alpha/beta ratio.

These results show that private information is attribute-specific rather
than concentrated in a single common feature. Consequently, suppressing
one highly ranked coordinate is unlikely to remove gender, age, and
identity information simultaneously.

\begin{table}[t]
\centering
\caption{Top ANOVA-ranked EEG features for private-attribute leakage.}
\label{tab:feature_responsibility}
\begin{tabular}{lllc}
\toprule
Attribute & Rank & Feature & $F$-value \\
\midrule
Gender
& 1
& Theta absolute power
& \textbf{75.135} \\

Gender
& 2
& Theta relative power
& 53.203 \\

Gender
& 3
& Alpha/Beta ratio
& 41.027 \\

\midrule

Age
& 1
& Beta relative power
& \textbf{42.683} \\

Age
& 2
& Temporal--frontal alpha asym.
& 41.877 \\

Age
& 3
& Left--right beta asym.
& 39.123 \\

\midrule

SubjectID
& 1
& Beta absolute power
& \textbf{116.644} \\

SubjectID
& 2
& Beta relative power
& 49.659 \\

SubjectID
& 3
& Alpha/Beta ratio
& 46.664 \\
\bottomrule
\end{tabular}
\end{table}

\paragraph{\textbf{Gaussian noise.}}

Table~\ref{tab:noise} reports task and private-attribute balanced
accuracy under Gaussian perturbation. The $\sigma=0$ configuration is
the original-feature baseline for this experiment.

We select $\sigma=0.30$ as the privacy-oriented operating point. At this
setting, task balanced accuracy changes only from 0.608 to 0.602, while
gender decreases from 0.691 to 0.633 and subject identity decreases from
0.718 to 0.594. However, age balanced accuracy increases from 0.322 to
0.415. The results indicate that isotropic perturbation affects private
attributes differently and does not provide consistent multi-attribute
suppression.

\begin{table}[t]
\centering
\caption{Balanced accuracy under Gaussian noise on standardized EEG
features.}
\label{tab:noise}
\begin{tabular}{lcccc}
\toprule
$\sigma$
& Task $\uparrow$
& Gender $\downarrow$
& Age $\downarrow$
& SubjectID $\downarrow$ \\
\midrule
0.00
& 0.608
& 0.691
& 0.322
& 0.718 \\

0.03
& 0.585
& 0.715
& 0.363
& 0.710 \\

0.05
& 0.611
& 0.688
& 0.339
& 0.703 \\

0.10
& 0.605
& 0.669
& 0.351
& 0.683 \\

0.20
& 0.617
& 0.659
& 0.409
& 0.642 \\

\textbf{0.30}
& \textbf{0.602}
& \textbf{0.633}
& \textbf{0.415}
& \textbf{0.594} \\
\bottomrule
\end{tabular}
\end{table}

\paragraph{\textbf{Feature masking}.}

Table~\ref{tab:masking} reports balanced accuracy after masking the
highest-ranked privacy-related EEG features. The ``None'' configuration
is the original-feature baseline for this experiment.

Top-8 is selected as the privacy-oriented masking configuration because
it provides the largest subject-identity reduction while preserving task
performance. SubjectID balanced accuracy decreases from 0.718 to 0.518,
while task balanced accuracy remains unchanged at 0.608. Gender decreases
modestly from 0.691 to 0.664, whereas age increases from 0.322 to 0.363.

Feature masking therefore provides a stronger reduction in identity
leakage than Gaussian perturbation in this experimental setting, but its
effect is not consistent across all private attributes.
\begin{table}[t]
\centering
\caption{Balanced accuracy under privacy-ranked feature masking. The
selected operating point is shown in bold.}
\label{tab:masking}
\begin{tabular}{lcccc}
\toprule
Masking
& Task $\uparrow$
& Gender $\downarrow$
& Age $\downarrow$
& SubjectID $\downarrow$ \\
\midrule
None
& 0.608
& 0.691
& 0.322
& 0.718 \\

Top-1
& 0.617
& 0.674
& 0.360
& 0.715 \\

Top-3
& 0.611
& 0.676
& 0.360
& 0.642 \\

Top-5
& 0.588
& 0.683
& 0.377
& 0.536 \\

\textbf{Top-8}
& \textbf{0.608}
& \textbf{0.664}
& \textbf{0.363}
& \textbf{0.518} \\
\bottomrule
\end{tabular}
\end{table}

\subsection{ \textbf{Privacy--Utility Performance.}}
Table~\ref{tab:overall_comparison} compares \sys with the original EEG
representation and the two feature-level baselines. Against the
original-feature baseline from the same partition, \sys preserves task
balanced accuracy at 0.781 vs 0.788, a reduction of 0.007
balanced-accuracy points.

At the same time, gender inference decreases from 0.858 to 0.563, age
inference from 0.789 to 0.467, and subject-identity inference from 0.692
to 0.206, corresponding to reductions of 0.295, 0.322, and 0.486
balanced-accuracy points. Subject-identity inference remains above the
closed-set chance level of 0.028, indicating that identity information is
substantially reduced but not completely removed.

The Gaussian-noise and feature-masking rows were obtained under a
different data partition and should be read against their own
original-feature baseline (0.608 task, 0.691 gender, 0.322 age, 0.718
subject identity) rather than against the first row of
Table~\ref{tab:overall_comparison}. Relative to that baseline, neither
lightweight defense reduces all three private attributes: both leave age
inference higher than it started. \sys instead reduces gender, age, and
subject identity within a single learned representation while maintaining
task performance close to the original features.

\begin{table*}[t]
\centering
\caption{Privacy--utility comparison across the evaluated methods. Each
privacy value is the strongest of the SVM, RF, and GB attackers, following
Eq.~\ref{eq:leakage}. For the original features, the strongest attacker is
RF for Task (0.788), GB for Gender (0.858), RF/GB tied for Age (0.789), and
RF for SubjectID (0.692).}
\label{tab:overall_comparison}
\begin{tabular}{lcccc}
\toprule
Method
& Task $\uparrow$
& Gender $\downarrow$
& Age $\downarrow$
& SubjectID $\downarrow$ \\
\midrule
Original features
& \textbf{0.788}
& 0.858
& 0.789
& 0.692 \\
Gaussian noise ($\sigma=0.30$)
& 0.602
& 0.633
& 0.415
& 0.594 \\
Feature masking (Top-8)
& 0.608
& 0.664
& \textbf{0.363}
& 0.518 \\
\sys
& 0.781
& \textbf{0.563}
& 0.467
& \textbf{0.206} \\
\bottomrule
\end{tabular}
\end{table*}

\paragraph{\textbf{Key takeaway.}}

The results demonstrate that EEG feature release preserves
substantial information about gender, age, and subject identity.
Gaussian noise and feature masking reduce selected forms of leakage, but
their effects vary across private attributes. \sys instead learns a
single representation that preserves the intended task while jointly
reducing all three evaluated private attributes.
\section{Discussion}
The results show that EEG feature release is not privacy-neutral.
The original feature representation supports the intended task, but it
also exposes demographic and identity information, and on this dataset it
exposes gender more reliably than the task itself. This is a systems
concern because compact feature release is widely treated as a safer
alternative to raw-signal transmission.

The two baselines illustrate the privacy--utility tension directly.
Gaussian noise and feature masking affect private attributes
differently. Gaussian perturbation reduces gender and identity at larger
noise levels but does not consistently reduce age inference. Feature
masking provides stronger identity reduction while largely preserving
task performance, but gender changes only modestly and age inference
does not decrease. These results show that fixed feature-level defenses
do not provide consistent multi-attribute privacy protection.Our \sys is more effective
because it learns a new representation optimized jointly for task utility
and multi-attribute suppression, and so acts on a subspace rather than on
the original feature coordinates.

The implication for wearable EEG pipelines is that privacy should be
treated as a property of the released representation rather than of the
raw signal alone. Protection must operate before compact features leave
the device.

\section{Limitations and Future Work}

Our evaluation has several limitations. First, we use a single dataset (EEGMAT) 
with 36 subjects in a controlled lab setting, which may not generalize to 
diverse populations or real-world deployment scenarios. Second, we evaluate 
closed-set subject identification over enrolled users; open-set or 
cross-session scenarios may present different privacy--utility tradeoffs. 
Third, our utility task is lab-based mental arithmetic, whereas real-world 
wearable EEG applications may involve different cognitive states, 
continuous monitoring, or multimodal sensing. Fourth, our attacker model 
assumes access to labeled auxiliary data from the same population; 
attackers with less or different side information may achieve weaker or 
stronger inference. Future work should validate \sys on diverse EEG 
datasets and deployment contexts, extend evaluation to open-set and 
cross-session identity scenarios, and explore additional privacy 
attributes such as health conditions, emotional states, or BCI experience. 
We also plan to deploy \sys on actual wearable devices to measure 
computational cost, latency, and energy consumption under real-time 
constraints, and to investigate whether the learned representation transfers 
effectively across tasks and populations.

\section{Conclusion}

This paper presented \sys, a learned representation framework for
privacy-preserving wearable EEG feature release. We first showed that
compact spectral and spatial EEG features are not privacy-neutral: the
same representation used for cognitive-state inference also exposes
gender, age, and subject identity. Feature-responsibility analysis
further showed that this leakage is structured across different spectral
and spatial feature groups. We evaluated Gaussian noise and feature
masking as lightweight defenses and found that their effects are
attribute-dependent, limiting their ability to provide consistent
multi-attribute privacy protection. \sys addresses this limitation by
learning a compact representation that preserves task-relevant
information while adversarially suppressing private attributes before
the representation is released. In validation, \sys reduces
private-attribute inference by up to 0.486 balanced-accuracy
points while maintaining task performance close to the original
feature representation. These findings suggest that practical
neuroprivacy for wearable EEG systems requires protection at the
representation level rather than relying only on raw-signal avoidance,
uniform perturbation, or fixed feature removal.

\bibliographystyle{ACM-Reference-Format}
\bibliography{references}

\end{document}